\documentclass[prb,showpacs,twocolumn,superscriptaddress]{revtex4}
\usepackage{graphicx,bm,amsmath,amssymb,psfrag}
\usepackage{color,xfrac,xcolor}
\newcommand{\Tr}[1]{\mbox{Tr} \left \{ #1  \right \}}
\begin{document}
\title{Gap formation in helical edge states with magnetic impurities}
\author{Simon Wozny}
\affiliation{School of Science and Engineering, Reykjavik University, Menntavegi 1, IS-101 Reykjavik, Iceland}
\affiliation{Fachbereich Physik, Universit\"{a}t Konstanz, D-78457 Konstanz, Germany}
\author{Karel Vyborny}
\affiliation{Institute of Physics, Academy of Science of the Czech Republic, Cukrovarnick\'{a} 10, Praha 6, Czech Republic}
\author{Wolfgang Belzig}
\affiliation{Fachbereich Physik, Universit\"{a}t Konstanz, D-78457 Konstanz, Germany}
\author{Sigurdur I.\ Erlingsson}
\email{sie@ru.is}
\affiliation{School of Science and Engineering, Reykjavik University, Menntavegi 1, IS-101 Reykjavik, Iceland}
  
\begin{abstract}
Helical edge states appear at the surface of two dimensional topological insulators and are characterized by spin up traveling in one direction and the spin down traveling in the opposite direction.  Such states are protected by time reversal symmetry and no backscattering due to scalar impurities can occur.  However, magnetic impurities break time reversal symmetry and lead to backscattering.  
Often their presence is unintentional, but in some cases they are introduced into the sample to open up gaps in the spectrum.
We investigate the influence of random impurities on helical edge states, specifically how the gap behaves in the realistic case of impurities having both a magnetic and a scalar component.  It turns out that for a {\em fixed} magnetic contribution the gap closes when either the scalar component, or Fermi velocity is increased.  
We compare diagrammatic techniques in the self-consistent Born approximation to numerical calculations which yields good agreement. For experimentally relevant parameters we find that even moderate scalar components can be quite detrimental for the gap formation.
\end{abstract}
\maketitle
\section{Introduction}
The transport properties observed in quantum Hall systems are determined by the presence of edge states, where both spin species travel in the same direction along the edges, so-called chiral edge states \cite{prange89:book}.  In quantum Hall systems time reversal symmetry (TRS) is broken (due to the applied magnetic field). A time-reversal symmetric version of the quantum Hall effect was proposed \cite{bernevig06:1757} in 2006 and a year later it was experimentally observed\cite{konig07:766}.  This quantum spin Hall effect is a generic property of two dimensional topological insulators (TI) \cite{hasan10:3045,qi11:1057}.
The edge states that appear in TIs are described by the one-dimensional (1D) massless Dirac equation \cite{hasan10:3045,qi11:1057}.  Graphene, which is described by the two-dimensional (2D) massless Dirac equation \cite{castroneto09:109}, can lead to peculiar transport properties such as Klein tunneling\cite{katsnelson06:620,beenakker08:1337}, i.e. perfect transmission through a potential barrier at normal incidence.  In 1D the electrons are always normal incident on the barrier and any type of {\em scalar} potential barrier will only lead to a phase factor in the wave function and thus always yield perfect transmission. Hence, the edge states are protected by TRS and backscattering is absent.

When a potential that breaks TRS is added, backscattering becomes possible.  Deviations from perfect transmission are observed in experiments and possible sources of such TRS breaking mechanisms have been proposed, e.g.\ inelastic scattering due to electron-electron interaction\cite{kainaris14:075118}, and tunnel-coupling to states in quantum dots\cite{vayrynen14:115309}, interaction with nuclear spins\cite{hsu17:081405} and interplay of Rashba spin-orbit coupling and magnetic impurities\cite{kimme16:081301}.  
\begin{figure}[t]
 \begin{center}
 \psfrag{l}{\huge $\ell$}
 \psfrag{N}{\huge $N_\mathrm{imp}$}
 \includegraphics[angle=0,width=0.35\textwidth]{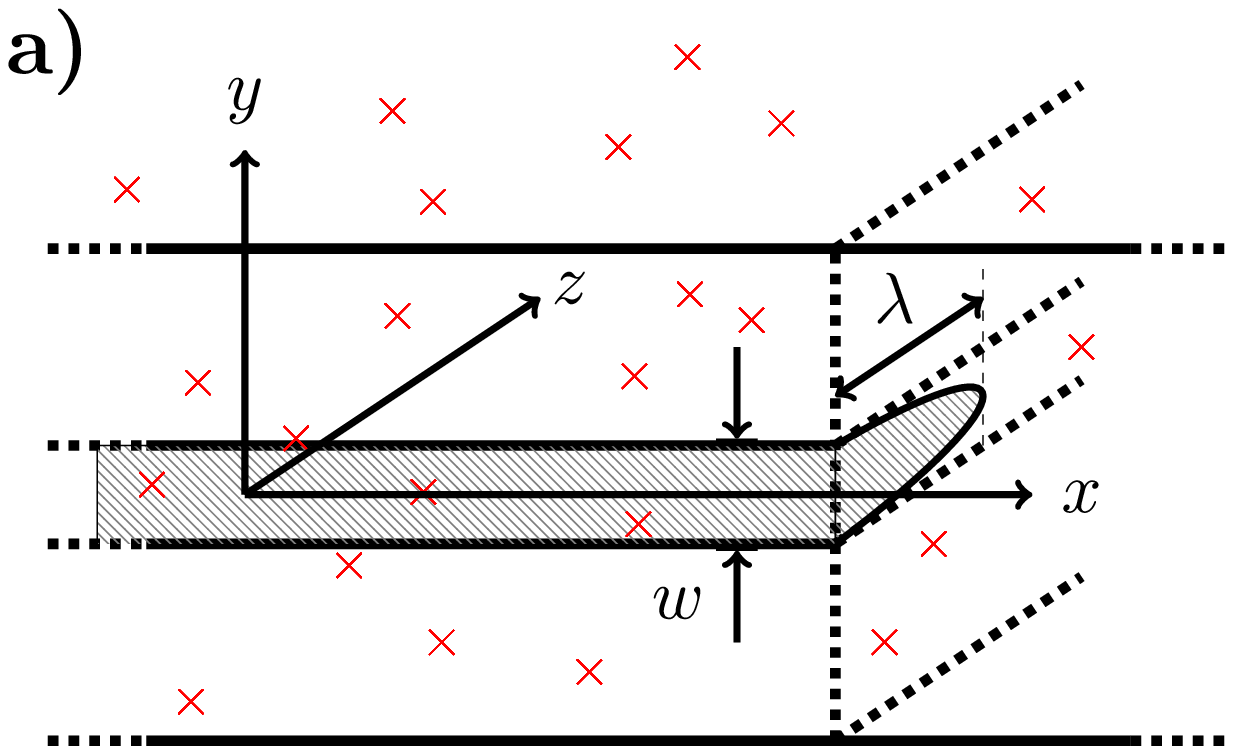}\\
  \vspace{0.4cm}
  \includegraphics[angle=0,width=0.45\textwidth]{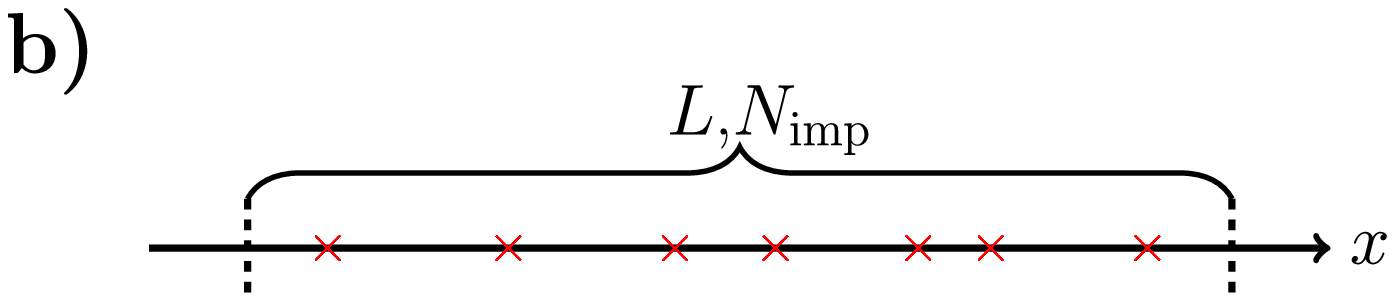}
  \caption{a) A schematic showing the edge state that appear when the bulk  TI sample is cut.  For sample widths much greater than $\lambda$ the edge states are decoupled.  The red crosses represent magnetic impurities distributed evenly throughout the bulk. b)  For the numerical calculations $N_\mathrm{imp}$ magnetic impurities are placed randomly along a length $L$.  The length $L$ is chosen such that tunneling through the impurity region is suppressed, i.e.\ a real gap can develop.}
  \label{fig:schematic}
 \end{center}
\end{figure}

In some cases the presence of magnetic impurities is even desired, e.g.\ for the quantum anomalous Hall effect (QAHE) \cite{qi11:1057}, and the sample is then intentionally doped.  
The helical edge states appear at the surface of a semiconductor heterostructure, such as HgTe/CdTe or InAs/GaSb systems.  In order to introduce magnetic impurities the material has to be doped at growth by magnetic atoms, e.g.\ Mn \cite{furdyna88:R29}.  However, it is not enough to have magnetic impurities present to achieve magnetic ordering of the impurity moments.  The ordering needs to be mediated by itinerant carriers \cite{jungwirth06:809}.  Unfortunately, early attempts to induce magnetic ordering in HgTe/CdTe failed\cite{liu08:146802}. 
More recently, it has been shown that the peculiar band structure of
InAs/GaSb and the singularities in the density of states (DOS) can lead to magnetic ordering of Mn moments \cite{wang14:147201} and the subsequent observation of the QAHE \cite{qi11:1057}.  This effect has also been observed\cite{chang13:167} in Cr doped (Bi,Se)Te.  In both of these systems the resulting magnetization is out-of-plane.  More recently, the observation of the QAHE has been proposed in systems, e.g.\ strained HgMnTe, with in-plane magnetization\cite{liu13:086802,ren16:085411}.  The goal of this paper is to investigate the behaviour of the helical edge states in the presence of such an in-plane magnetization.
When the magnetic moments of the impurities align they lead to a (random) magnetic field, but it will still have a non-zero average.  The resulting net magnetic field will open up a gap in the energy spectrum, which could be observed spectroscopically in the DOS \cite{qi11:1057}.

The detection of a gap in the DOS is frequently performed in superconductor heterostructures. In particular, the smearing of the
superconducting gap due to the proximity effect has been addressed in
this way in normal metal-superconductor heterostructures
\cite{belzig:96,gueron:96,scheer:01,lesueur:08}. Furthermore, the
connection to Andreev states was established by tunneling into carbon
nanotubes or graphene \cite{pillet:10,bretheau:17}. More recently, tunneling into semiconducting nanowires in proximity to a superconductor has been used to find evidence for Majorana modes \cite{mourik:12,deng:16} with a possible applications in topological quantum computing.

The magnetic impurity atoms will not only lead to local magnetic moments.  They affect the electrostatic environment around them and can thus lead to both magnetic and electric potentials localized around the impurity positions \cite{furdyna88:R29}.
Earlier work considered the effects of magnetic impurities with potential parts on the local DOS in 3D TIs\cite{black-schaffer15:201411}.  Here we will consider the DOS of the helical edge states in the presence of magnetic doping giving rise to in-plane magnetization, and how it is affected by the scalar potential contribution.  Due to the high density of impurities the DOS is obtained using the usual averaging techniques.  We will use both diagrammatic techniques and direct numerical calculations (Secs.\ \ref{sec:diagrammatics} and \ref{sec:numerics}) to study the influence of the scalar contribution on the gap formation. 
As a result we find that the simultaneous presence of a scalar and magnetic contributions the magnetic gap is first reduced for moderate scalar potential and then closes for sufficiently strong scattering (Sec.\ \ref{sec:results}).  The gap is also reduced  as the Fermi velocity is increased, although more slowly as compared to the scalar potential increase.

\section{Model and methods}
\label{sec:model}
Helical edge states appear on boundaries of 2D
TIs\cite{bernevig06:1757,qi11:1057}, which can be formed in quantum
well semiconductor heterostructures as schematically shown in
Fig.~\ref{fig:schematic}a).  The helical states circulate along the
edges of the system, just like an edge state in the quantum Hall effect,
but the former are helical with the spin locked to the momentum
due to strong spin-orbit interaction. If the sample width is much
greater than the edge state width $\lambda$, the edges can be
considered isolated and they are then described by the Hamiltonian (assuming the edge is along the x-direction)
\begin{equation}
 H_0=\hbar v_F k_x  \sigma_z ,
\label{eq:H0}
\end{equation}
where $v_F$ is the Fermi velocity of the edge modes, $k_x$ the wave vector and $\sigma_z$ the third Pauli matrix in spin-space.  The two
counter-propagating spin modes are protected by TRS,
i.e.\ backscattering can only occur if a term that breaks TRS, e.g.\ 
magnetic field or magnetic impurities, is added.

To study the influence of impurities in this context, it is
instructive to look at how the edge states are embedded in the 3D
structure.  A schematic of a typical setup including impurities is shown in
Fig.\ \ref{fig:schematic}a). The schematic
represents a system formed in a quantum well as in the case of
HgTe/CdTe heterostructures\cite{konig07:766} or at the interface of
two materials as for InAs/GaSb based systems\cite{liu08:236601}.
Only those impurities which are within reach of the edge state wave
function, contribute to the effective 1D model, as explained in
App.\ \ref{app:3Dto1D}.  The magnetic impurities are assumed to be
aligned along a specific axis $\bm{M}$, as it is considered in
Ref.\ \onlinecite{wang14:147201}.  For our purpose, it is not
important which specific direction is chosen, as long as
$\bm{M}$ has components other than the $z$-component, so we choose
$\bm{M}||x-$axis. The effective one-dimensional Hamiltonian we
are going to consider is then 
\begin{align}
 H&=\hbar v_F k_x  \sigma_z+\sum_j (V\sigma_0+M \sigma_x)\delta(x-x_j),
 \label{eq:H}\\
 	&=H_0+U_V(x)+U_M(x) \nonumber
\end{align}
where we have added\cite{black-schaffer15:201411} a scalar
contribution $V \sigma_0$ and a magnetic contribution $M \sigma_x$ to
the full Hamiltonian. Note that the sum over $j$ represents random
impurity positions along the
$x$-axis whose average separation is $d$ (which is inherited from the
3D impurity distribution), as shown in Fig.\ \ref{fig:schematic}b).
For the diagrammatic methods considered in
Sec.\ \ref{sec:diagrammatics} the length $L$ is considered to be
infinite while $n_\mathrm{imp}=N_\mathrm{imp}/L$ is kept constant.
For the finite size numerics in Sec.\ \ref{sec:numerics}, $L$ is finite
but large enough to allow the gap to develop.

\subsection{Diagrammatics and impurity averaging}
\label{sec:diagrammatics}
\begin{figure}[h]
 \psfrag{BA=}{\large $\Sigma_{BA}=$}
 \begin{center}
  \includegraphics[angle=0,width=0.45 \textwidth]{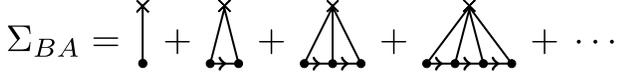}
  \caption{Diagrammatic representation of the full Born approximation.}
  \label{fig:selfBA}
 \end{center}
\end{figure}

In order to calculate the DOS of the system, we start with the Green's
function (from now on, we write $k_x=k$) 
\begin{eqnarray}
 G^R(k,E)=\frac{1}{ G^{-1}_0(k)-\Sigma(E)},
\label{eq:Gfull}
\end{eqnarray}
where $ G_0(k)=[E- H_0(k)]^{-1}$ is the unperturbed Green's
function in the momentum domain (corresponding to $ H_0$ in
Eq.\ (\ref{eq:H0})) and $ \Sigma(E)$ is the irreducible self-energy
due to the impurity contribution after averaging over 
impurity configurations. We can now introduce a diagrammatic notation
representing a scatterer by a cross, a single scattering process by a
line, and a propagation by an arrowed line\cite{doniach98:book}. In
the full Born approximation\cite{bruus04:book} (BA), also known as t-matrix approximation, we sum over an infinite
set of diagrams as shown in Fig. \ref{fig:selfBA}.

The series can be summed up using the t-Matrix method and the resulting
equation, after introducing $x_M=\frac{M}{2\hbar v_F}$ and
$V_M=\frac{V}{M}$, for the irreducible self-energy reads
\begin{eqnarray}
   \Sigma_\mathrm{BA} &=& n_\mathrm{imp}M
  \frac{ \sigma_x+ V_M-i x_M ( 1- V_M^2 )}{(1+ iV_M x_M )^2+x_M^2}.
\label{eq:SE_BA}
\end{eqnarray}
In the limit of $x_M\ll 1$ and $V_M \ll 1$ the self-energy reduces to
$ \Sigma_\mathrm{BA}=n_\mathrm{imp} M \sigma_x$, corresponding
to a homogeneous magnetization.  Note the selfenergy does not depend
on $E$ and, as we sHall see later, will not lead to a fully developed gap in the
density of states. Furthermore Eq.\ (\ref{eq:SE_BA}) is only linear in
$n_\mathrm{imp}$ (the dilute limit).

\begin{figure}[h]
 \psfrag{BA=}{\large \!\!\!\!\!\!\!\!$\Sigma_{SCBA}=$}
 \begin{center}
  \includegraphics[angle=0,width=0.45 \textwidth]{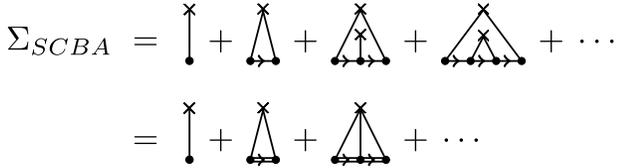}
  \caption{The diagrams constituting the self-consistent Born approximation.  Note that the second line is the same as in Fig.\ \ref{fig:selfBA} with full Green's functions instead of bare ones.}
  \label{fig:selfSCBA}
 \end{center}
\end{figure}

A convenient way to improve the BA, i.e.\ to include more diagrams, is
the self-consistent Born approximation (SCBA) which diagrammatically
has the form shown in Fig.\ \ref{fig:selfSCBA} and could be called here self-consistent t-matrix approximation.

In the Born approximation, see Fig.\ \ref{fig:selfBA}, the electrons
only scatter off one impurity (one cross), but in the SCBA an infinite
number of crosses appears thereby improving the approximation.  However, only
non-crossing diagrams are included in the SCBA.  The self-energy in
the SCBA is obtained by iterating the self consistency equation for
the self-energy using the t-Matrix method: Defining $
t(E)= U+ U\sum_k G^R(E,k) t(E)$ the self-energy can be
expressed as $ \Sigma_{SCBA}(E)=n_{imp} t(E)$ and the
equations are closed by the Dyson equation $ G^{R-1}(k,E)=
G_0^{-1}(k,E)+\Sigma_{SCBA}(E)$.

Once the self-energy has been found the retarded version of the Green's function in
Eq.\ (\ref{eq:Gfull}) can be used to find the DOS
\begin{equation}
 \mathcal{D}(E)= -\frac{1}{\pi} \int \frac{dk}{2\pi}\mathrm{tr}\mathrm{Im}\{  G^R(k,E)\}.
\end{equation}
The DOS will then show a gap around zero energy, and in the following
sections we will consider how the gap is affected by the scalar
part $V$ of the impurity potential.

\subsection{Numerical procedure}
\label{sec:numerics}
In the absence of the impurities the spectrum consists of two linear
dispersion modes, right-- and left--movers resulting from the $\pm 1$
eigenvalues of the Pauli matrix $\sigma_z$.  It turns out the
potential part of the impurities only leads to a phase factor and the
Green's function 
%
\begin{equation}
 [E- ( -i \hbar v_F \partial_x \sigma_z+U_V(x))]g^R_V(x,x';E)=\delta(x-x'),
 \label{eq:g0EOM}
\end{equation}
where $U_V$ denotes the scalar part of the impurity potential in
Eq.\ (\ref{eq:H0}).  The Green's function in Eq.\ (\ref{eq:g0EOM}) can
be found exactly, resulting in
\begin{eqnarray}
 g^R_V(x,x';E)
  &=& e^{-i\frac{\sigma_z V}{\hbar v_F}\sum_n(\theta(x-x_n)-\theta(x'-x_n))  }\nonumber \\
  &&\times g^R_0(x-x';E) 
  \label{eq:gRV}
\end{eqnarray}
where $g_0^R(x,E)$ is the Green's function for the homogeneous system 
\begin{eqnarray}
 g^R_0(x,E)&=&\frac{-i}{2\hbar v_F } \left [
 e^{i\frac{Ex}{\hbar v_F}}\theta(x)(1+\sigma_z) \right. \nonumber  \\
 & &+\left. 
 e^{-i\frac{Ex}{\hbar v_F}}(1-\theta(x))(1-\sigma_z)\right].
 \label{eq:gR0}
\end{eqnarray}
The Heaviside functions $\theta$ reflect the helical properties,
i.e.\ 'spin up' $(1 \quad 0)^T$ is a right--mover, and 'spin down' $(0
\quad 1)^T$ is a left--mover. When $x\rightarrow 0$ in the argument
of $\theta$ in Eq.\ (\ref{eq:gR0}), the Heaviside functions
need to be calculated in the weak sense, such that $ \lim_{x \rightarrow 0}\theta(x)=1/2$.

When both magnetic and potential impurities are considered, one can
start from the equation of motion for the Green's function of the full
impurity system 
\begin{eqnarray}
 G^R(x,x')\!\!&=&\!\!g^R_V(x,x') \nonumber+ \int d\bar{x}g^R_V(x,\bar{x})U_M(\bar x) G^R(\bar{x},x') \nonumber \\
 &=&\!\!g^R_V(x,x')+ M \sum_n g^R_V(x,x_n) \sigma_x G^R(x_n,x').
 \label{eq:Geom} 
\end{eqnarray}
In Eq.\ (\ref{eq:Geom}), we have omitted the $E$ arguments in $g^R_V$
and $G^R$ for the sake of brevity.
Multiplying Eq.\ (\ref{eq:Geom}) with $\hbar v_F$ we obtain a
dimensionless equation for corresponding $\tilde{G}^R$ and
$\tilde{g}^R_V$.  Evaluating $\tilde{G}^R$ at positions $x=x_l$ and
$x'=x_m$, we otbain a set of $N$ linear equations,
\begin{eqnarray}
[\tilde{G}^R]_{l,m}&=&[\tilde{g}^R_V]_{m,l}+ \frac{M}{\hbar v_F}\sum_{n=1}^N [\tilde{g}^R_V]_{l,n} \sigma_x[\tilde{G}^R]_{n,m} \nonumber \\
\label{eq:Gteom} 
\end{eqnarray}
From the $N \times N$ problem above, one can construct an equation for
matrices of dimension $2 N \times 2N$
\begin{equation}
A\tilde{G}^R=\tilde{g}^R_V
\end{equation}
where the matrix $A$ is defined as
\begin{equation}
 A_{l,m}=\delta_{l,m}\sigma_0-\frac{M}{\hbar v_F}\tilde{g}^R_V(x_l,x_m) \sigma_x.
\end{equation}
What is left is to invert matrix $A$ to obtain
\begin{equation}
 \tilde{G}^R(x_l,x_m,E)= [\tilde{G}^R]_{l,m}= [A^{-1}\tilde{g}^R_V]_{l,m}.
\label{eq:invAgVR}
\end{equation}
It should be noted that although the calculations involve
$\tilde{G}^R(x_l,x_m,E)$ evaluated at discrete points, the method
avoids fermion doubling since Eq.\ (\ref{eq:Gteom}) is a discretized
integral equation leading to {\em non-local} coupling of the lattice points.   
From this we can calculate the DOS at a given position
\begin{equation}
  \mathrm{\mathcal{D}}(x_l,E)=
  -\frac{1}{\pi} \Tr{\mathrm{Im} [ G^R(x_l,x_l,E)]}.
 \label{eq:IDOSnumx}
\end{equation}
Since our goal is to compare to results obtained using diagrammatic
methods (corresponding to infinite system size) we use the
DOS at the center of impurity region, i.e\ we
define $x_c$ as the impurity site closest to $L/2$.  The value
$\mathrm{\mathcal{D}}(x_c,E)$ is then calculated by averaging over
different impurity configurations.

\section{Results and discussion}
\label{sec:results}
The strength of magnetic impurities is characterized by the parameter $E_M$ which is typically of the order $0.5-1.0$\,eV for Mn doped III-V material\cite{wang14:147201}.  
Using the cation density of the host material $n_0=N_c /a^3$, where $N_c$ is number of cations per unit cell and $a$ is the lattice constant, $E_M$ can be related to $\bm{M}_\mathrm{3D}$, see details in App.\ \ref{app:3Dto1D}.
For InAs and HgTe $N_c=4$ unit cells and $a \approx 6.1$\,$\AA$ and $a \approx 6.5$\,$\AA$, respectively. 
We can translate this to a numerical value for the 1D magnetization
\begin{equation}
 |\bm{M}|=\frac{E_Ma^3}{N_c d^2}=\frac{E_Ma x_\mathrm{eff}^{2/3}}{N_c},
\end{equation}
where $x_\mathrm{eff}$ denoted the fraction of magnetic impurities $n_\mathrm{3D}=x_\mathrm{eff} n_0$, see App.\ \ref{app:3Dto1D}. 
Taking the values for HgTe and Mn and assuming $x_\mathrm{eff}=1\%$ gives $|\bm{M}|=0.07$\,eV$\AA$.  This quantity, along with $\hbar v_F= 3.0$\,eV$\AA$ results in $x_M \approx 0.01$ for $E_M=0.5$\,eV.  For InAs/GaSb based systems $\hbar v_F \approx 0.3$\,eV$\AA$, resulting in $x_M \approx 0.1$, assuming other parameters describing the magnetic impurities remain relatively unchanged.  
Hence, in our numerical calculations we choose values of $x_M\lesssim 0.1$, which can be realistically achieved in experiments.
\begin{figure}[h]
 \begin{center}
  \includegraphics[angle=-0,width=0.5\textwidth]{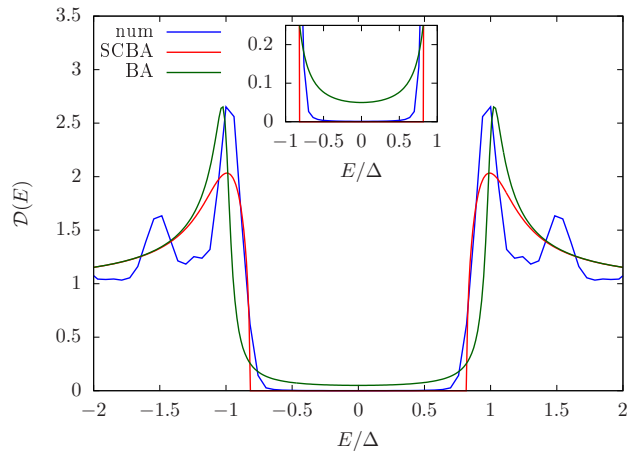}
  \caption{The DOS $\mathcal{D}$ as a function of energy with $x_M=0.05$ in the absence of scaler scattering ($V=0$) for different approximation discussed in the text.  The DOS obtained numerically by impurity averaging shows oscillations related to finite size effects, which are absent for the SCBA and BA.  The inset shows a zoom of the three curves into the gap region, note that the BA fails to show a hard gap.}
  \label{fig:BA_SCBAcomparison}
 \end{center}
\end{figure}

The three traces in Fig.\ \ref{fig:BA_SCBAcomparison} are the results for the Born approximation (BA) [green], the self-consistent BA (SCBA) [red], and the numerical impurity averaging [blue].  In the numerics a finite amount of impurities, typically ranging from $N=40$ to $80$, is used, so that the effective system size is finite.  This introduces two effects not present in the BA or SCBA.  First, there are oscillations related interference terms that add contributions proportional to  Re$\{e^{iE L/\hbar v_F}\}$.  Second, when the energy approaches the gap edges the barrier penetration becomes significant leading to some states penetrating the gap, as can be seen in the inset.  
A homogeneous magnetic field, which induces a term $\Delta\sigma_x$ added to $H_0$ in Eq.\ (\ref{eq:H0}), results
in a gap with edges at energies $E= \pm \Delta$.  But for the magnetic impurities the gap is slightly reduced due to fluctuations in the impurity configuration.

Now we turn our attention to the diagrammatic results.  Once the self-energy is known, see Eq.\ (\ref{eq:SE_BA}), the DOS $\mathrm{\mathcal{D}}(E)$ can be obtained.  As is clear from the inset in Fig.\ \ref{fig:BA_SCBAcomparison} the self-energy in the BA is always finite and the density does not vanish in the gap.  In fact it can be shown that $\mathrm{\mathcal{D}_{BA}}(E=0) \propto x_M$. 
The BA results can be improved in the SCBA, which can be obtained using an iteration loop to reach a self-consistent Green's function \cite{bruus04:book}.
In the SCBA an infinite set of new diagrams is included leading to a self-energy that becomes energy dependent in and around the gap.  Note that the SCBA does not include crossing diagrams.  The SCBA curve shows good agreement with the BA outside the gap and inside the gap the numerical curve and SCBA show good agreement although the finite size effects in the numerics preclude an exact match.
\begin{figure}[h]
 \begin{center}
  \includegraphics[angle=0,width=0.48\textwidth]{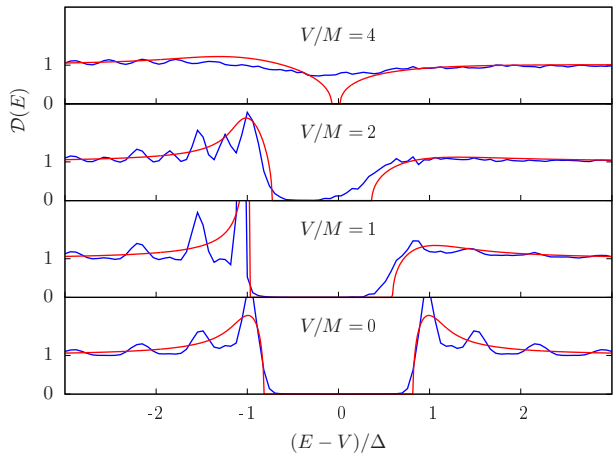}
  \caption{The density of states $\mathcal{D}$ (within SCBA [red] and numerical approach [blue]) shown for $x_M=0.05$ and for four different values $V/M=0, 1, 2$, and $4$.  The energy is shifted by $V$ to align the gap centers.  The gap closes at around $V/M=4$.}
    \label{fig:stack_xM0.05}
 \end{center}
\end{figure}

Next we want to investigate how the gap changes when a scalar part is added to the impurity potential.  The scalar part is modeled using the term $V \sigma_0$ where $V$ is measured relative to $\Delta$.  In Fig.\ \ref{fig:stack_xM0.05} the density of states is shown for four different values of $V/M=0, 1, 2$ and $4$.  As the value of $V$ is increased the gap can clearly be seen to shrink, both in the numerics and the SCBA.  Also, the center of the gap shifts to higher energies.  This is a result of the real part of the self-energy being proportional to $V$ to first approximation.  In Fig.\ \ref{fig:stack_xM0.05} the energy is shifted by $V$ to reflect this.   There is some discrepancy between the numerics and the SCBA on the upper gap edge which is most likely due to finite size issues that result in increased tunneling tails into the gap.  However, it can not be excluded that the non-crossing diagrams might affect the self-energy around the gap edges which the SCBA does not properly describe.  But the overall trend is clear: the scalar part $V$ rapidly leads to a closing of the gap at around $V=4M$.  The numerical curve shows that the gap has vanished but the SCBA gap closes only at $V \approx 4.3M$, not shown here. 

Having established how the scalar contribution to the impurity potential leads to a closing of the gap we look at different values of $x_M=0.1$ and $x_M=0.02$.  Experimentally this could be achieved by changing the impurity concentration $n_i$, or considering different host materials with different values of $v_F$.
\begin{figure}[h]
 \begin{center}
  \includegraphics[angle=0,width=0.5\textwidth]{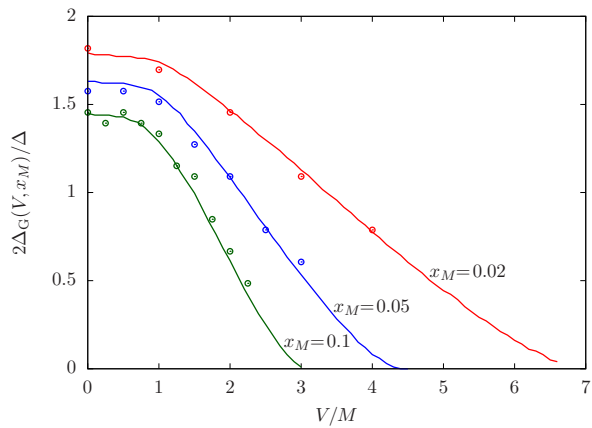}
  \caption{The gap size plotted as function of $V$ for three different values of $x_M=0.02, 0.05$ and 0.1.  The solid lines show the SCBA result and the open circles are the numerical results.  The SCBA shows that the gap cap closes at $V/M\approx 3, 4.3$ and $6.7$ for $x_M=0.1, 0.05$, and $0.02$, respectively.}
  \label{fig:gapVxM}
 \end{center}
\end{figure}
The behavior of the gap size as a function of $V$ for different values of $x_M=0.02, 0.05$ and $0.1$ is shown in Fig. \ref{fig:gapVxM}.
As can be seen in Fig.\ \ref{fig:stack_xM0.05} for $V/M=4$ the gap has closed for the numerical results due to finite size effects and the SCBA shows a closing at slightly higher values of $V$.  The SCBA shows that the gap cap closes at $V/M\approx 3, 4.3$ and $6.7$ for $x_M=0.1, 0.05$ and $0.02$, respectively.  In the numerics the gap edges are defined where $\mathcal{D}(E_\pm)=1/2$.
The results in Fig.\ \ref{fig:gapVxM} clearly show that scalar contribution in the impurity potential can have quite a detrimental effect on the gap size.  This has consequences for creating gaps in 2D TIs using magnetic impurities.  The magnetic material and/or the host TI material should be chosen to have as low values of $V$ as possible, but having 
high value of $v_F$ (which leads to low values of $x_M$ for a fixed $M$) can mitigate the gap reduction.
\section{Summary}
We have shown that random magnetically ordered impurities can open up a gap for helical edge states, but the inevitable scalar contribution to the impurity potential suppresses the gap considerably.  
Assuming {\em fixed} values of $M$ and $n_\mathrm{imp}$ the gap size is influenced by two parameters: $V$ and $\hbar v_F$, or in terms of dimensionless quantities $V/M$ and $x_M$.
The gap is reduced by either increasing $V$ or decreasing $v_F$,  as can be seen in Fig.\ \ref{fig:gapVxM}.
This behavior was observed in both the diagrammatic approach using the SCBA and a full numerical approach with impurity averaging.  
This has experimental consequences and the magnetic dopants used should have as low a scalar contribution as possible.

\begin{acknowledgments}
WB acknowledges useful discussions with G. Rastelli and R. Klees and the hospitality of the Center for Quantum Spintronics (QuSpin) at NTNU Trondheim, Norway.
This work was supported by the Reykjavik University Research Fund, Erasmus$+$ and the DFG through RA 2810/1.
\end{acknowledgments}

\appendix
\section{3D to 1D impuritity properties}
\label{app:3Dto1D}
The material hosting the edge channels is doped with magnetic impurities of density $n_\mathrm{3D}$.  Note that the impurities are embedded in the (3D) host material. Assuming short range scattering, the magnetic impurities can be described by
\begin{equation}
 U_\mathrm{3D}(x,y,z)=\sum_{x_j,y_j,z_j} \bm{M}_\mathrm{3D}\cdot\bm{\sigma}\delta(x-x_j)\delta(y-y_j)\delta(z-z_j).
 \label{eq:U3D}
\end{equation}
The goal here is to extract the influence of the impurities in the region of the edge states.  This is accomplished by using the edge state wave function  $\chi(y,z)$, reflecting the thickness $w$ of the two-dimensional system and the extension of the edge state into the bulk $\lambda$, see Fig.\ \ref{fig:schematic}a). The influence of the magnetic impurities on the one-dimensional helical edge states is obtained by projecting Eq.\ (\ref{eq:U3D}) onto the edge state
\begin{eqnarray}
 U(x)&=&\int dy dz |\chi(y,z)|^2 U_\mathrm{3D}(x,y,z) \nonumber \\
 &=&\sum_{x_j} \bm{M}\cdot \bm{\sigma} \delta(x-x_j).
 \label{eq:U1D}
\end{eqnarray}
In the last step we introduced
\begin{equation}
 \bm{M}=\sum_{y_j,z_j}\bm{M}_\mathrm{3D} |\chi(y_j,z_j)|^2 \approx d^{-2}\bm{M}_\mathrm{3D},
 \label{eq:M1D}
\end{equation}
where $d \sim n_\mathrm{3D}^{-1/3}$ is the (average) distance between impurities, and we replaced the sum over $(y_n,z_n)$ with an integral $\int dydz$, and assumed that $|\chi \rangle$ is normalized.

\end{document}